\documentclass[reprint,amsmath,amssymb,aps]{revtex4-1}
\usepackage{xcolor}
\usepackage{graphicx}
\usepackage{subfigure}
\usepackage{amsfonts}
\usepackage{bm}
\usepackage{dcolumn}
\usepackage{color}
\usepackage{mathtools}

\usepackage{xcolor}
\colorlet{shadecolor}{orange!15}

\usepackage{mathrsfs}
\usepackage{dsfont}
\usepackage{wasysym}

\usepackage[skins,theorems]{tcolorbox}
\tcbset{highlight math style={enhanced,
  colframe=red,colback=white,arc=0pt,boxrule=1pt}}

\usepackage{amsmath}

\newcommand{\Pt}{\text{P}}

\begin{document}

\newcommand*{\cofrac}[2]{%
  {%
    \rlap{$\dfrac{1}{\phantom{#1}}$}%
    \genfrac{}{}{0pt}{0}{}{#1-#2}%
  }%
}
\title{Chain decay and rates disorder in the totally asymmetric simple exclusion process}
\author{Y. Ibrahim$^{1,2}$, J. Dorignac$^{1}$, F. Geniet$^{1}$, C. Chevalier$^{1}$,  J. C. Walter$^{1}$, N-O. Walliser$^{1}$,  A. Parmeggiani$^{1}$, J. Palmeri$^{1}$}

 \affiliation{$^{1}$Laboratoire Charles Coulomb, UMR5221 CNRS-UM, Universite de Montpellier, Place Eugene Bataillon, 34095 Montpellier Cedex 5, France \\
$^{2}$Department of Physics, Umaru Musa Yar'adua University, P.M.B. 2218 Katsina, Nigeria}

\date{\today}

\begin{abstract}
%
%
We theoretically study the Totally Asymmetric Exclusion Process (TASEP) with quenched jumping rates disorder and finite lifetime chain. TASEP is widely used to model the translation of messenger RNAs by Ribosomes in protein synthesis. Since the exact solution of the TASEP model is analytically and computationally intractable for biologically relevant systems parameters, the canonical Mean-Field (MF) approaches of solving coupled non-linear differential equations is also computational expensive for the scale of relevant biological data analysis. In this article, we provide alternative approach to computing the MF steady state solution via a computationally efficient system of non-linear algebraic equations. We further outline a framework for including correlations progressively via the exact solution of small size TASEP system. Leading order approximation in the biologically relevant entry rate limited regime shows remarkable agreement with the full Monte-Carlo simulation result for a wide range of system parameter space. These results could be of importance to the kinetic rates inference in Ribo-Seq data analysis and other related problems.
\end{abstract}
\maketitle

\section{Introduction}
The totally asymmetric exclusion process (TASEP) is a paradigmatic model of non-equilibrium statistical mechanics \cite{derrida1993exact,parmeggiani2003phase,derrida2007non}. It is a simple model that captures the essential features of many non-equilibrium systems, such as traffic flow \cite{szavits2019accurate,ciandrini2013ribosome}, diffusion in biological membranes, and protein synthesis.
The TASEP is a one-dimensional lattice model with particles that can hop to their nearest neighbor site in the forward direction, but not in the backward direction. The particles are hard-core, meaning that no two particles can occupy the same site at the same time (excluded volume interaction). The TASEP is driven by a difference in the particle density between the two ends of the lattice \cite{derrida2004asymmetric}.
TASEP has been studied extensively using both analytical and numerical methods \cite{derrida1993exact,derrida1998exactly,neri2011totally}. It has been shown that the TASEP exhibits a variety of non-equilibrium phenomena, such as phase transitions and jamming.
The TASEP can be used to model protein synthesis by considering the ribosomes as particles on a lattice, where the lattice sites represent codons on the messenger RNA (mRNA) \cite{macdonald1969concerning}. The ribosomes can hop to the next codon on the mRNA in the forward direction, but not backwards. 
The TASEP has been used to model a variety of aspects of protein synthesis, such as the ribosome flow rate (current), the distribution of ribosomes on the mRNA, and the effect of collisions/jamming on protein synthesis \cite{greulich2008phase}.

Here, we study the TASEP with quenched jump rates disorder and finite degradation rate of the chain. We give an alternative framework of approximating the system correlations and outlined computationally iterative solution method thereby circumventing the coupled non-linear differential equations.
%
%
%
%
%
%
%
%

The article is organized as follows. In the next section, we outline the TASEP model and introduce notations. The main results are presented in section III while we discuss the results implication and conclude in section IV.

\begin{table*}
\centering
\begin{tabular}{| c | l |  c |   }
\hline 
\ Dimensionless ratio  \    &  \  Description   \   \\
\hline \hline
 &  \\
$\Omega/w_{min} \in [0,\infty)$  & \ Score quantifying the chain degradation rate relative to the weakest jump rate in the bulk   \\
 &   \\
$\alpha / w_{min} \in  (0,\infty) $   & \  This quantifies the onset of phase transition out of the Low Density phass \\
 &   \\
 $\Delta \equiv 1 - {w_{min}}/{w_{max}}  \in  [0,1) $ & \  The jump rates disorder distribution width quantifies the chances of occurrence of a slow bond \\
 & \  right after a fast bond and thereby inducing a traffic jam  \\
 &   \\
\hline 
\hline
\end{tabular} 
\caption{Relevant dimensionless ratios for the five kinetic rates: $\alpha, w_{min}, w_{max}, \beta$ and the chain degradation rate $\Omega$. Note that unless otherwise stated, we assume the exit rate $\beta$ to be non-limiting (i.e. $\beta \gg \alpha, w_{min}, w_{max}, \Omega$).} 
 \label{system:parameters}
 \end{table*}

\section{The TASEP model}
We consider a one-dimensional lattice with a total of $N$ lattice sites. We further denote the presence of a particle at a lattice site $j$ with $\sigma_j=1$ and its absence with $\sigma_j=0$. A particle, with $\ell$ lattice sites long footprint, jumps from site one site to another on the lattice with the following dynamical rules:
\begin{itemize}
%
\item[(a.)] A particle enter the lattice {at a rate $\alpha$} with its \emph{trailing edge} on the first site $(j=1)$, provided the first $\ell$ sites are empty ($\sigma_1 = \cdots = \sigma_{\ell} = 0$).
\item[(b.)] Particle advance from site $j$ to site $j+1$ {with a rate $w_j$}, provided the $j+1$ through $j+\ell$ sites are empty ($\sigma_{j+1} =  \cdots = \sigma_{j+\ell}=0$).
\item[(c.)] From the lattice site $j=N-\ell+1$, particle {exit} the lattice incrementally and {unhindered} {with a rate $w_j$}.
\item[(d.)] When the trailing edge of the particle is at site $N$, the particle exits with a rate $\beta$.
\end{itemize}
%

%
\begin{widetext}
\onecolumngrid

\begin{figure}
\includegraphics[scale=.5]{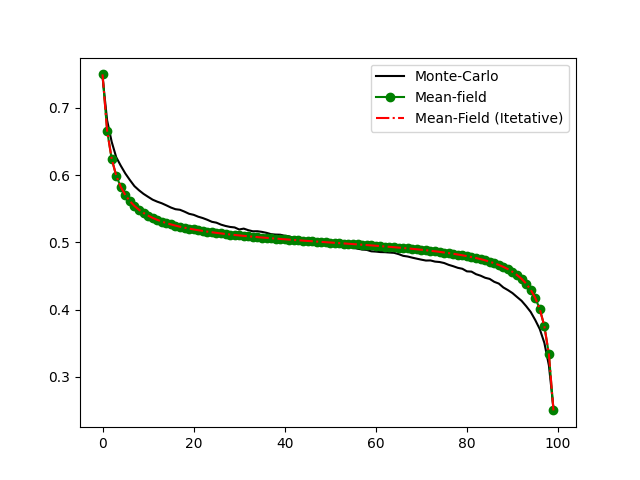}
\includegraphics[scale=.5]{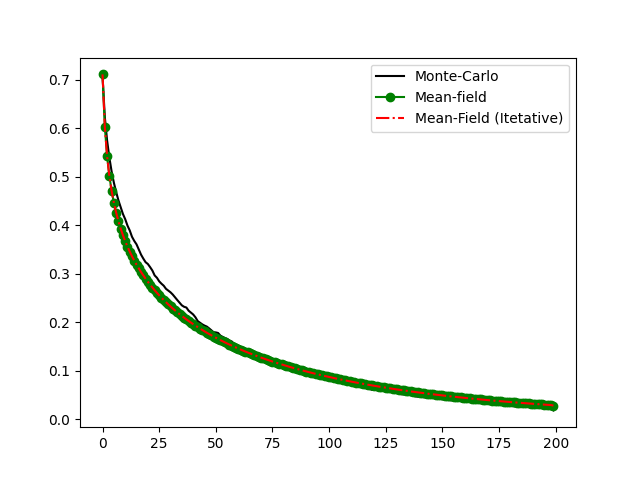}
\caption{Density profiles \textbf{(left)} $\Omega=0$ and \textbf{(right)} $\Omega = 0.01$. The other parameters remain constant: $\alpha=\beta = w_{max} = w_{min} = 1$.}
\label{fig:density_profiles0}
\end{figure}

\twocolumngrid
\end{widetext}

\subsection{Master equation}
The probability, $P(\mathcal{C},t)$, of finding the lattice in a configuration $\mathcal{C} \equiv \left(\sigma_1,\cdots,\sigma_N \right)$ at time $t$, evolve according to the master equation \cite{derrida1993exact,parmeggiani2003phase}
\begin{align}
\frac{d}{dt} \widetilde{P}\left( \mathcal{C},t \right)  &=  \sum_{ \mathcal{C}' } W_{ \mathcal{C} \leftarrow \mathcal{C}' } \,  \widetilde{P}\left( \mathcal{C}' \right)  -  \sum_{ \mathcal{C}' } W_{ \mathcal{C}' \leftarrow \mathcal{C} }  \, \widetilde{P}\left( \mathcal{C} \right)  \  ,  \label{eqn:master:equation}
\end{align}
where $W_{\mathcal{C} \leftarrow \mathcal{C}'}$ is the transition rate from state $\mathcal{C}'$ to state $\mathcal{C}$ and $ \widetilde{P}\left( \mathcal{C},t\right)$ is the probability of finding the system in configuration $\mathcal{C} \equiv \left(\sigma_1,\cdots,\sigma_N \right)$ at time $t$. $\sigma_j=1$ if site $j$ is occupied by the \emph{trailing} edge of the particle and $\sigma_j=0$ if empty.

Throughout this article, we assume that the chain has a constant degradation rate $\Omega$ and an exponential age distribution, $ \Omega \exp{\left(-\Omega \, t \right)} $,
such that the age averaged probabilities are \cite{valleriani2010turnover,szavits2020dynamics}
\begin{equation}
P\left( \mathcal{C} , \Omega \right)  \equiv \int_0^{\infty}  \widetilde{P}\left(\mathcal{C},t \right) \,\Omega \exp{\left(-\Omega \, t \right)}  \, dt  \  . 
\end{equation}
Therefore, the chain's age averaged master equation now reads
\begin{align}
&\Omega P\left( \mathcal{C},\Omega \right) - \Omega P(\mathcal{C}, t=0)  = \nonumber \\
& \quad \qquad \sum_{ \mathcal{C}' } W_{ \mathcal{C} \leftarrow \mathcal{C}' } \, P\left( \mathcal{C}', \Omega \right)  -  \sum_{ \mathcal{C}' } W_{ \mathcal{C}' \leftarrow \mathcal{C} } \,  P\left( \mathcal{C}, \Omega \right)  \  , \label{eqn:master:equation:age:averaged}
\end{align}
where we set the initial condition $P(\mathcal{C}, t=0)$ to an empty chain, i.e. $P(\mathcal{C}, t=0) = P\left( \sigma_1=\cdots=\sigma_N=0 \right)$. 

Solving for the $P\left( \mathcal{C},\Omega\right)$ will allow us to obtain moments of the particle site occupation, $\sigma_j$, such as the marginal probability $ \left<  \sigma_j \right> \equiv \sum_{\mathcal{C}} \sigma_j P\left( \mathcal{C}, \Omega \right) $.

The configuration space, $\mathcal{C}$, grows exponentially with the system size, $N$. Therefore, solving for $P\left(\mathcal{C} \right)$ becomes computationally expensive for physically relevant system sizes \citep{ciandrini2013ribosome} 


The marginals $\left< \sigma_j \right>$ and $\left<\sigma_j \sigma_{j+\ell} \right>$ satisfy the following system of algebraic equations \citep{parmeggiani2004totally,ciandrini2013ribosome}
\begin{align}
0 &=  \ \overbrace{  w_{j-1} \Big( \left< \sigma_{j-1} \right> - \left< \sigma_{j-1} \sigma_{j-1+\ell}\right>  \Big) }^{\text{flux into site $j$}}  \nonumber \\  & \qquad \quad   - \underbrace{  w_j \Big( \left< \sigma_j \right> - \left< \sigma_{j} \sigma_{j+\ell}  \right> \Big)  \quad -  \Omega \, \left< \sigma_j \right> }_{\text{flux out of site $j$}}  \  , \label{eqn:first:of:marginals:equation}
\end{align}
for $1 \leqslant j \leqslant N$, with fixed boundary conditions $ \sigma_0 =1$ and $ \sigma_{N+1} =  \sigma_{N+2} = \cdots = \sigma_{N+\ell} = 0 $ for the left and right boundaries respectively. Henceforth, we use interchangeably the notations $w_0 \equiv \alpha$ for the entry (initiation) rate and $w_N \equiv \beta$ for the exit (termination) rate.
\section{Results}
\begin{widetext}
\onecolumngrid
\begin{figure}
\includegraphics[scale=.5]{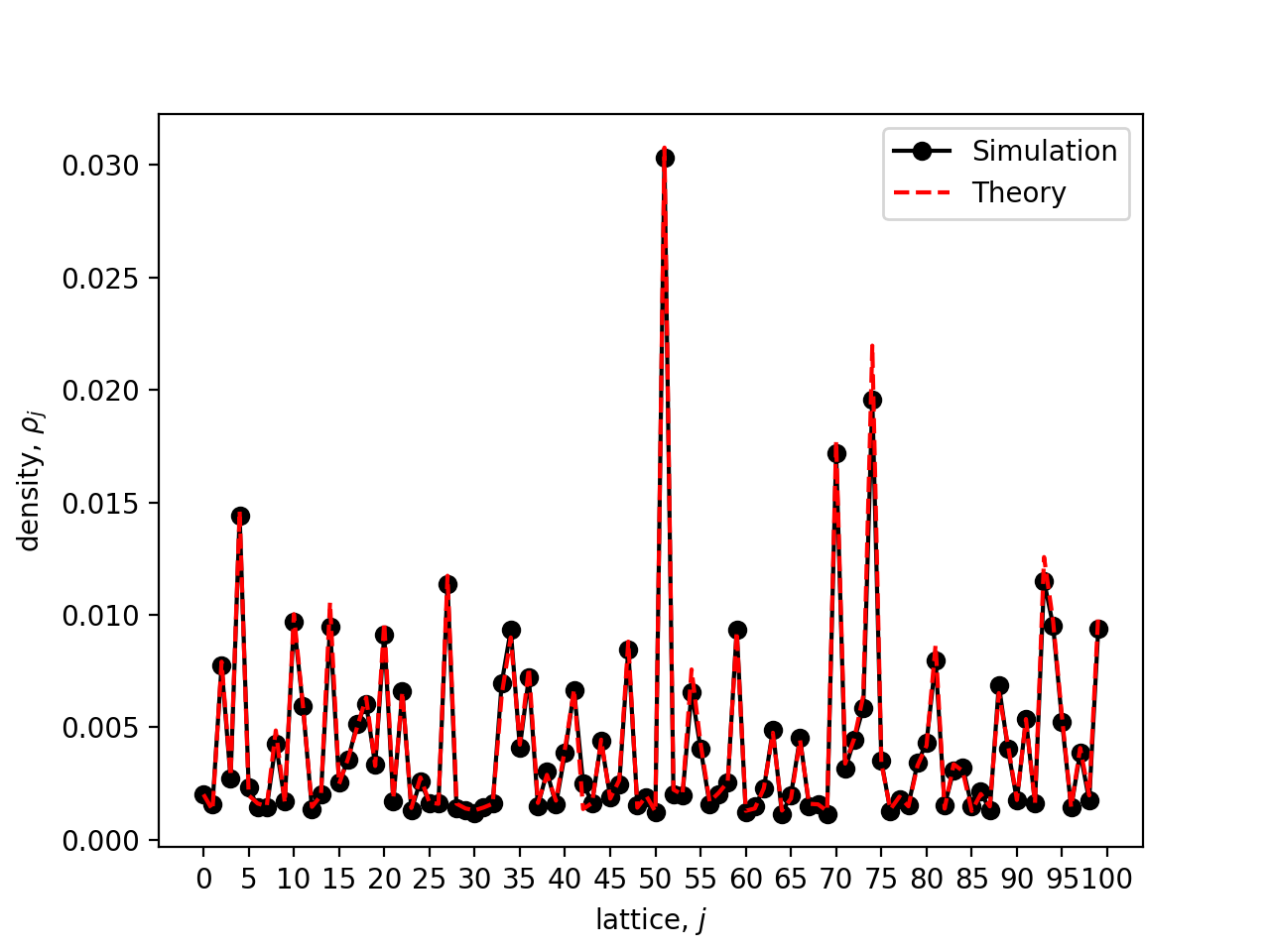}
\includegraphics[scale=.5]{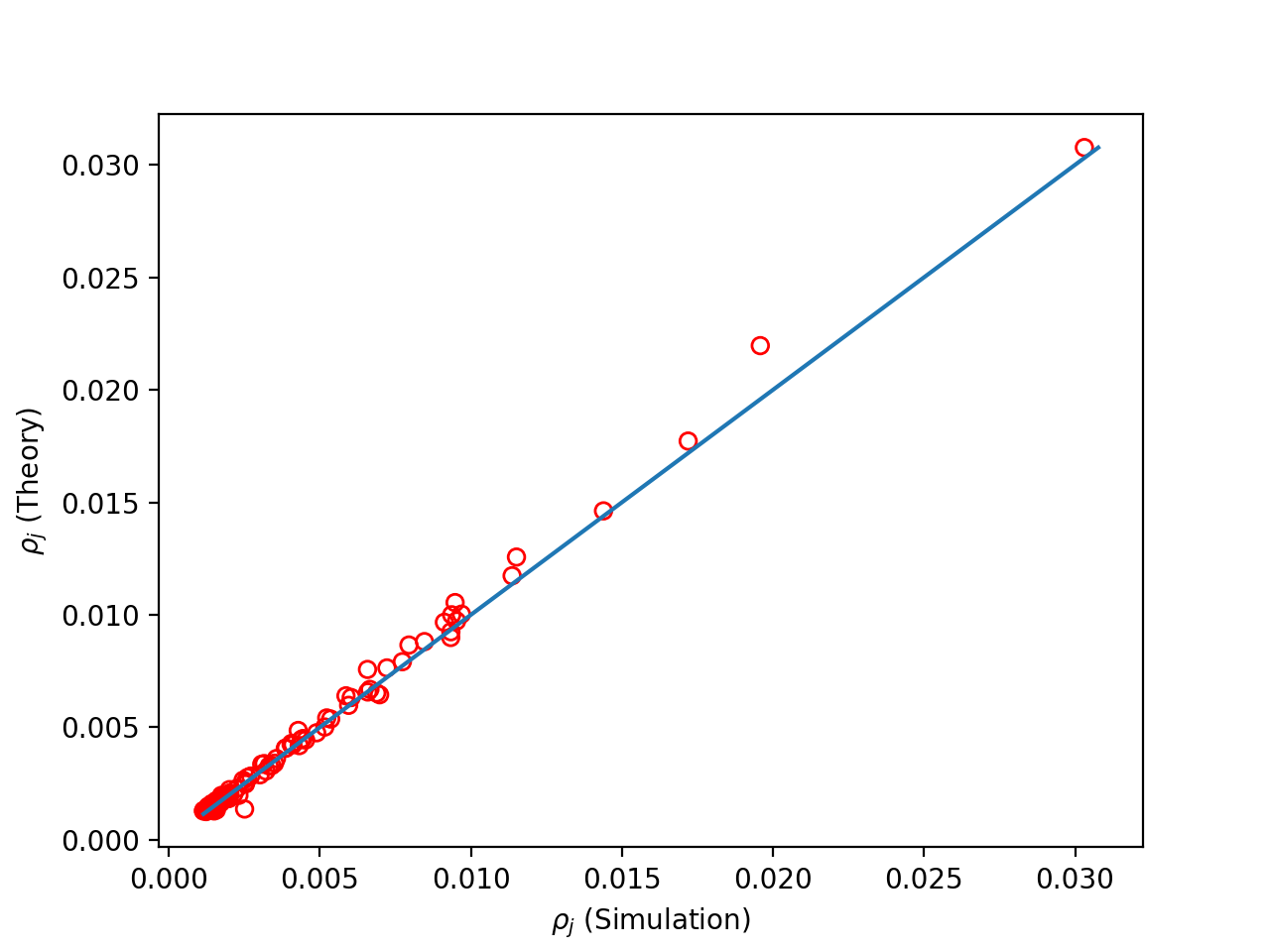}
\includegraphics[scale=.5]{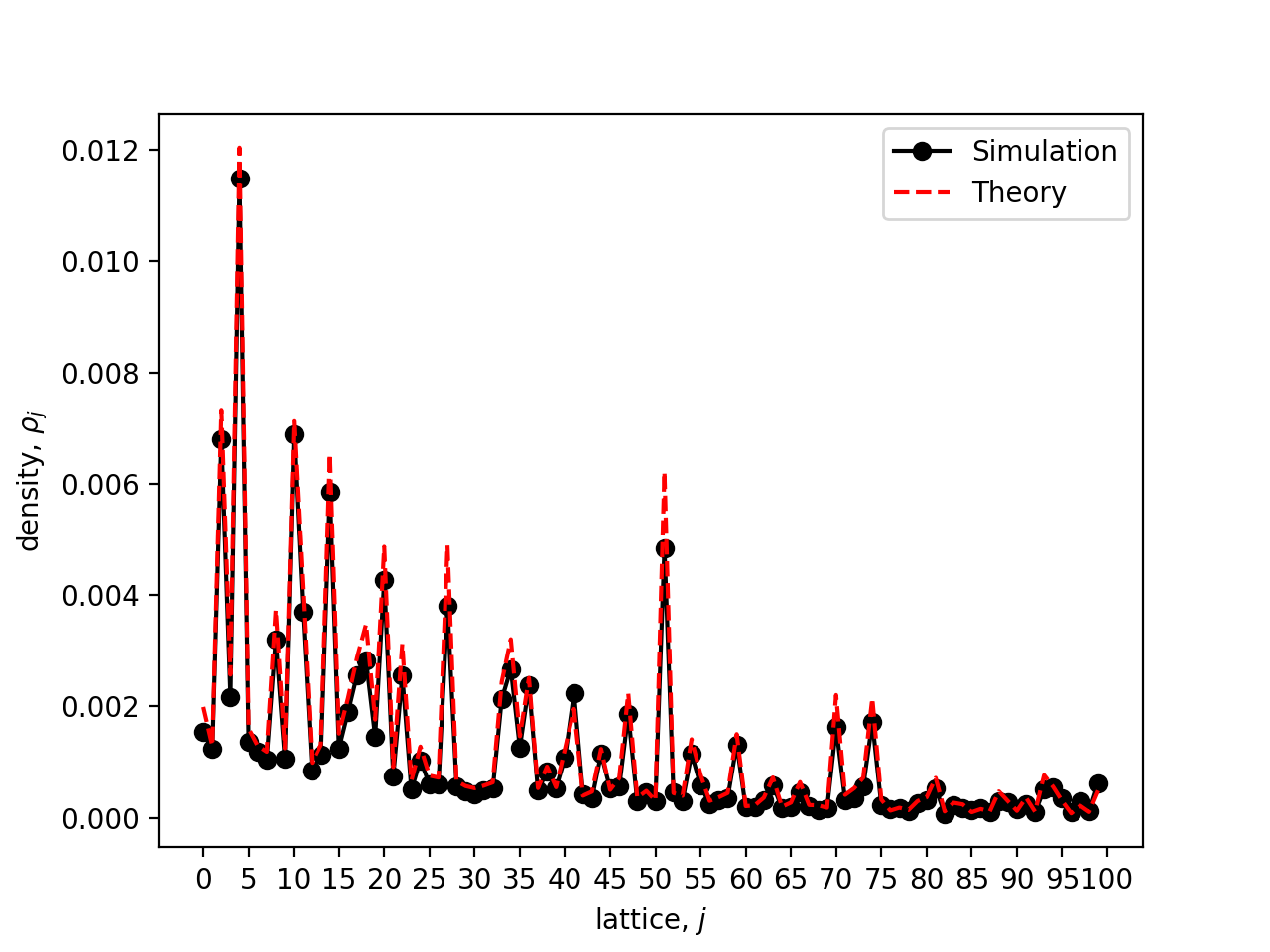}
\includegraphics[scale=.5]{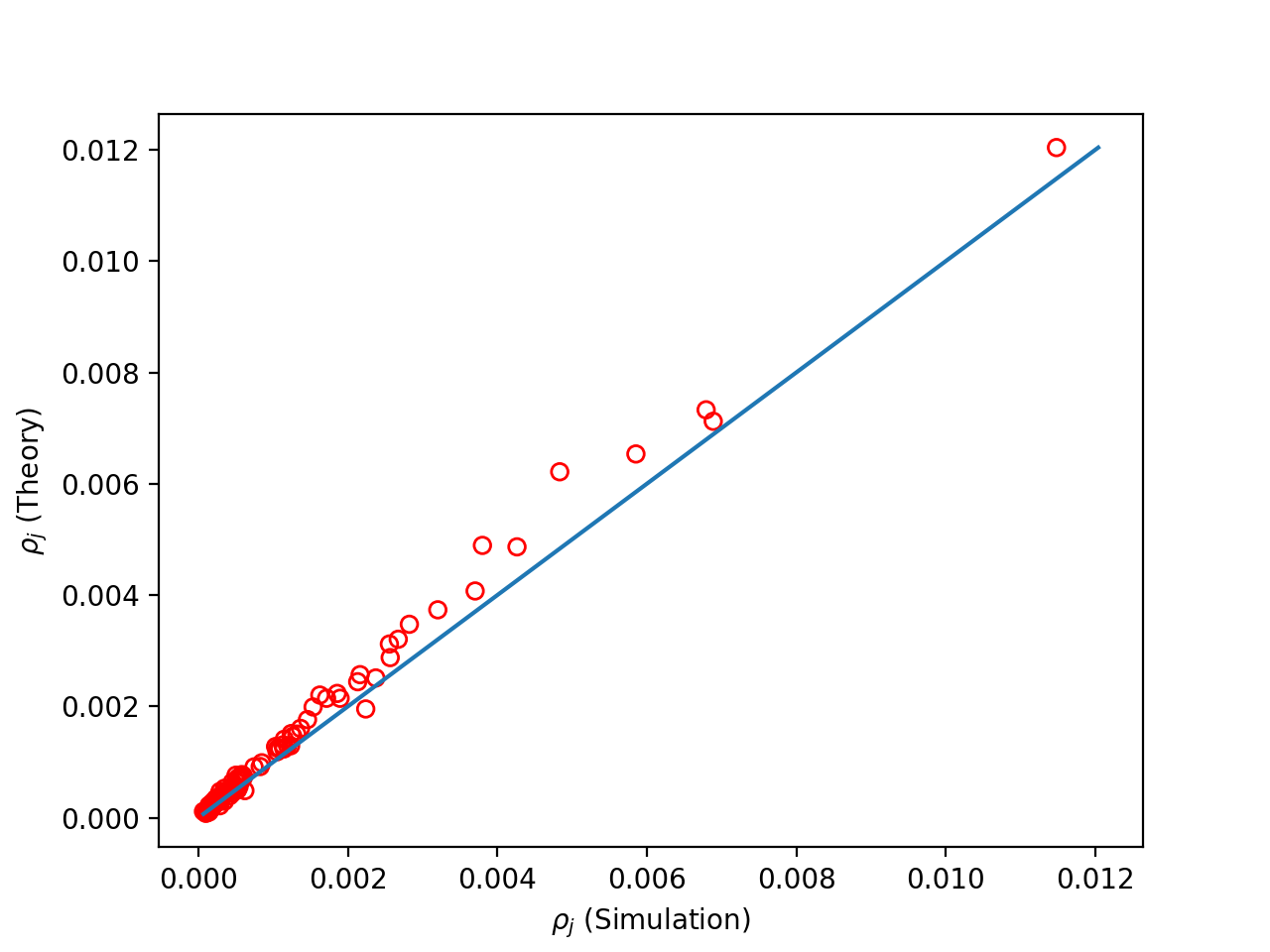}
\caption{(Chain decay, $\Omega/w_{min} > 0$) Densities, $\rho_j$, and scatter plots for sample sequence drawn from $w_j \in [3,80]$, $\alpha=0.1$, $\beta=30$, with $\ell=9$, and $N=100$ total lattice sites. (top left) $\rho_j$ plot for $\Omega=0$. The red dashed line is Eqn. (\ref{eqn:first:order:solution}) fit of the Monte Carlo simulation. (top right) Scatter plot of data points in the top left figure. (bottom left) $\rho_j$ plot for $\Omega=0.5$. The dashed red line is the Eqn. (\ref{eqn:first:order:solution}) fit. (bottom right) Scatter plot of data points in the bottom left figure. }
\label{fig:density_1}
\end{figure}
\twocolumngrid
\end{widetext}
\subsection{Correlations modeling}
To calculate the particle densities $\left< \sigma_j \right> \equiv \rho_j$ from eqn. (\ref{eqn:first:of:marginals:equation}), we require what we call the \emph{collision} marginals $ \left< \sigma_{j-1}\sigma_{j+\ell -1} \right>$ and $ \left< \sigma_j \sigma_{j+\ell} \right> $ to close the system of equations. The canonical approach is to build a hierarchy of equations of marginals involving larger number of sites (i.e. the BBGKY hierarchy) and truncating the system at a desired level of accuracy. With this approach, increasing accuracy of the approximation comes with linear increase in the number of equations to be solved. Secondly, the effectiveness of the approach depends on the quality of the approximation of the truncating correlation function.

Here, we take an entirely different approach, we model the two-site correlation marginal $\langle \sigma_j \sigma_{j+\ell} \rangle$ with the exact solution of few sites TASEP system. For $N=2$ and $N=3$ sites, the analytic calculation is tractable (see the Appendix). For larger system sizes, fast linear solvers could be used to solve for the correlation.


\subsubsection{$\ell = 1$ particles}
Hence, the marginal probability that a site $j+1$ is empty is $\left( 1 -  \rho_{j+1}\right)$ while the marginal probability that site $j$ is occupied is simply $\rho_j$. However, the probability of two particles colliding (i.e. site $j+1$ is occupied given that site $j$ is also occupied) is approximated by simple mutual independence of the sites \cite{macdonald1969concerning}
\begin{equation}
 \left< \sigma_{j} \sigma_{j+1} \right>  \rightarrow  \rho_j \, \rho_{j+1}   \  .
\end{equation}
This is the canonical Mean-Field (MF) approximation. 
It's important to note that we expect weak contributions from these terms in the entry limited regime of the dynamics.
This MF approximation greatly simplifies the problem from an exponentially large system of $\sim 2^N$ linear equations (eqns. \ref{eqn:first:of:marginals:equation}) to a relatively small system of $N$ non-linear coupled algebraic equations:
\begin{equation}
\Omega \, \rho_j  = w_{j-1} \rho_{j-1} \left( 1- \rho_{j} \right)  - w_j \rho_j \left( 1 - \rho_{j + 1} \right) \  ,   \label{eqn:mean:field:canonical}
\end{equation}
for $1\leqslant j \leqslant N$, 
and at the boundary, $\rho_0=1$, while $\rho_{N} = 0$.

\subsubsection{$\ell \geqslant 2$ (extended) particles}
In similar manner, for extended particles that have cover more a single site, the marginal probability that a site $j+\ell$ is empty is $\left( 1 - \sum_{s=1}^{\ell} \rho_{j+s} \right)$ while the marginal probability that site $j$ is occupied is simply $\rho_j$. McDonalds's et. al. \cite{macdonald1969concerning} provided an accurate approximation of the collision marginal
\begin{equation}
 \left< \sigma_{j} \sigma_{j+1} \right>  \rightarrow  \frac{\rho_j \, \rho_{j+\ell} }{ \left( 1 - \sum_{s=1}^{\ell} \rho_{j+s} \right) + \rho_{j+\ell} }   \  .
\end{equation}
There are $\ell$ modes of occupying each lattice site, $j$, and the empty mode \cite{macdonald1969concerning}. 
Essentially, the joint probability, $ \left< \sigma_{j} \sigma_{j+1} \right>$, is replaced by a product of the marginal probability of particle occupying site $j$ and the conditional probability that site $j+1$ is also occupied while site $j$ is either empty or occupied by the leading edge of the particle.

Substituting the correlation $\left< \sigma_{j} \sigma_{j+1} \right> $ in eqn. (\ref{eqn:first:of:marginals:equation}), it follows that
\begin{equation}
\Omega \, \rho_j  = w_{j-1} \rho_{j-1} \left( 1- R_{j-1+\ell} \right)  - w_j \rho_j \left( 1 - R_{j + \ell} \right) \  ,   \label{eqn:mean:field:canonical:extended}
\end{equation}
for $1 \leqslant j \leqslant N-\ell$ and 
$$
R_{j+\ell} = \frac{\rho_{j+\ell}}{\left( 1 - \sum_{s=1}^{\ell} \rho_{j+s}  \right) + \rho_{j+\ell}} ,   \quad 1\leqslant j \leqslant N-\ell
$$
while at the boundaries $\rho_0 = 1$ and $R_{N-\ell+1}=\cdots = R_{N+\ell}=0$.

\subsubsection{Iterative method of solution}
Instead of solving the $N$ system of coupled differential equations, the age-averaged densities could be directly computed from $N$ coupled algebraic equations:
\begin{equation}
\rho_j = \frac{w_{j-1}\rho_{j-1} \left( 1 - R_{j-1+\ell} + \rho_j \right) }{\Omega + w_{j-1} \rho_{j-1} + w_j \left( 1 - R_{j+\ell} \right) }
\end{equation} 
where starting with an appropriate initial guess, say $\rho_j = \mbox{min}(\alpha, \{ w_j \}, \beta)/(1+\ell)$, the system will converge to the solution after few iterations. This provide a computationally efficient way to compute the densities and other statistics of interest.

\subsection{Entry limited (low density) expansion}
\begin{figure}
\includegraphics[scale=.5]{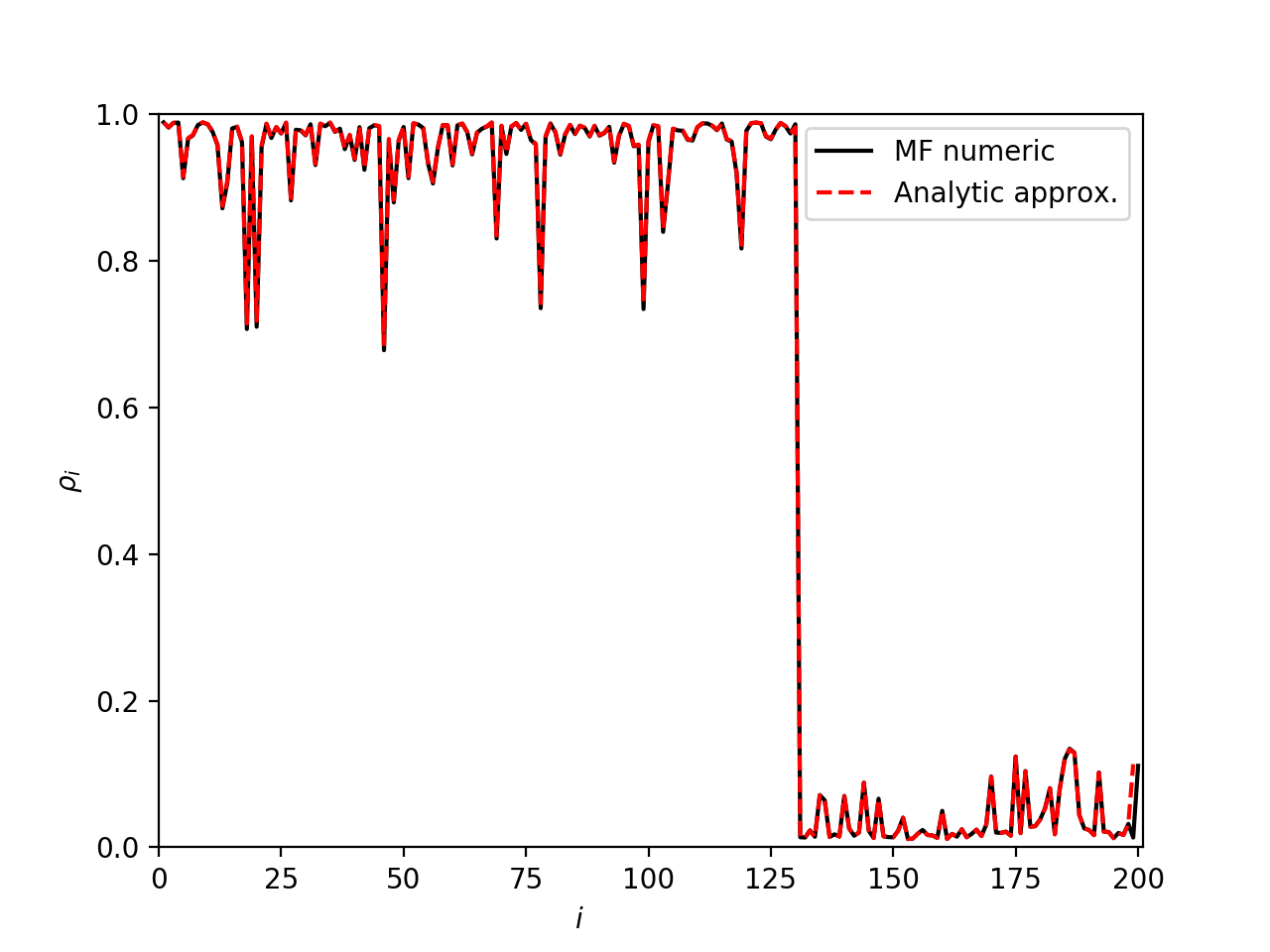}
\caption{Phase separation for $\alpha/w_{min} \gg {1}/{2}$ while $\beta/w_{min} \gg 1$, $\Omega = 0$ and $\Delta = 0.9$.}
\label{fig:phase:separation}
\end{figure}
Of special interest is the entry limited regime (or Low Density phase) in the parameter space due to its relevance to the biological process of ribosome translation of mRNA \cite{}. From eqn. (\ref{eqn:mean:field:canonical}), and expanding the densities in the small parameter $\epsilon = \alpha/(\Omega + w_1)$, we obtain the first order contribution in the small parameter, $\epsilon$, 
\begin{equation}
\rho_j \  =   \  \frac{\alpha}{w_j}   \prod_{n=1}^{j} \left( \frac{w_{n}}{\Omega + w_{n} } \right) \  +  \    \mathcal{O}\left( \epsilon^2 \right)  \   ,  \label{eqn:first:order:solution}
\end{equation}
for $1\leqslant j \leqslant N-1$ (see the Appendix for details). Interestingly, this first order contribution captures the essential physics in this entry limited regime, showing excellent quantitative agreement with the full Monte Carlo simulations (see Figs. \ref{fig:density_1}, and \ref{fig:density_2}) for a broad range of values. Expansion to the second order in $\epsilon$ could be found in the Appendix. For brevity, we keep here only the first order contribution.
%

\subsubsection{Reduced throughput}
To quantify the effects of both the chain decay and jumping rates disorder, we define a (dimensionless) throughput ratio (TR) quantity as, $\text{TR} = J_{\text{out}}/J_{\text{in}}$, where $J_{\text{in}}$ and $J_{\text{out}}$ are the entry and exit fluxes of particles respectively. In the low density limit, ($\alpha/w_{min} \ll 1$),
\begin{equation}
\text{TR} = \prod_{n=1}^N \left( \frac{w_n}{\Omega + w_n} \right) \  .
\end{equation}
Notably, for stable chains with infinite lifetime ($\Omega=0$), $\text{TR}=1$. While for chains with finite lifetime ($\Omega > 0$), the TASEP has reduced throughput, $0 < \text{TR} <1$. For uniform hopping rates, $w_n = \bar{w}$, $TR = \exp\left( - N \log\left( 1 + {\Omega}/{\bar{w}} \right)  \right)$. This implies that the particle current $J$ decays exponentially with the chain length $N$. For long chains, throughput is also exponentially sensitive to the ratio $\Omega/w_{min}$, and less obviously to the disorder score $\Delta = 1 - w_{min}/w_{max}$ (see Table \ref{table:dimensionless:ratios}).

\begin{widetext}
\onecolumngrid

\begin{figure}
\includegraphics[scale=.5]{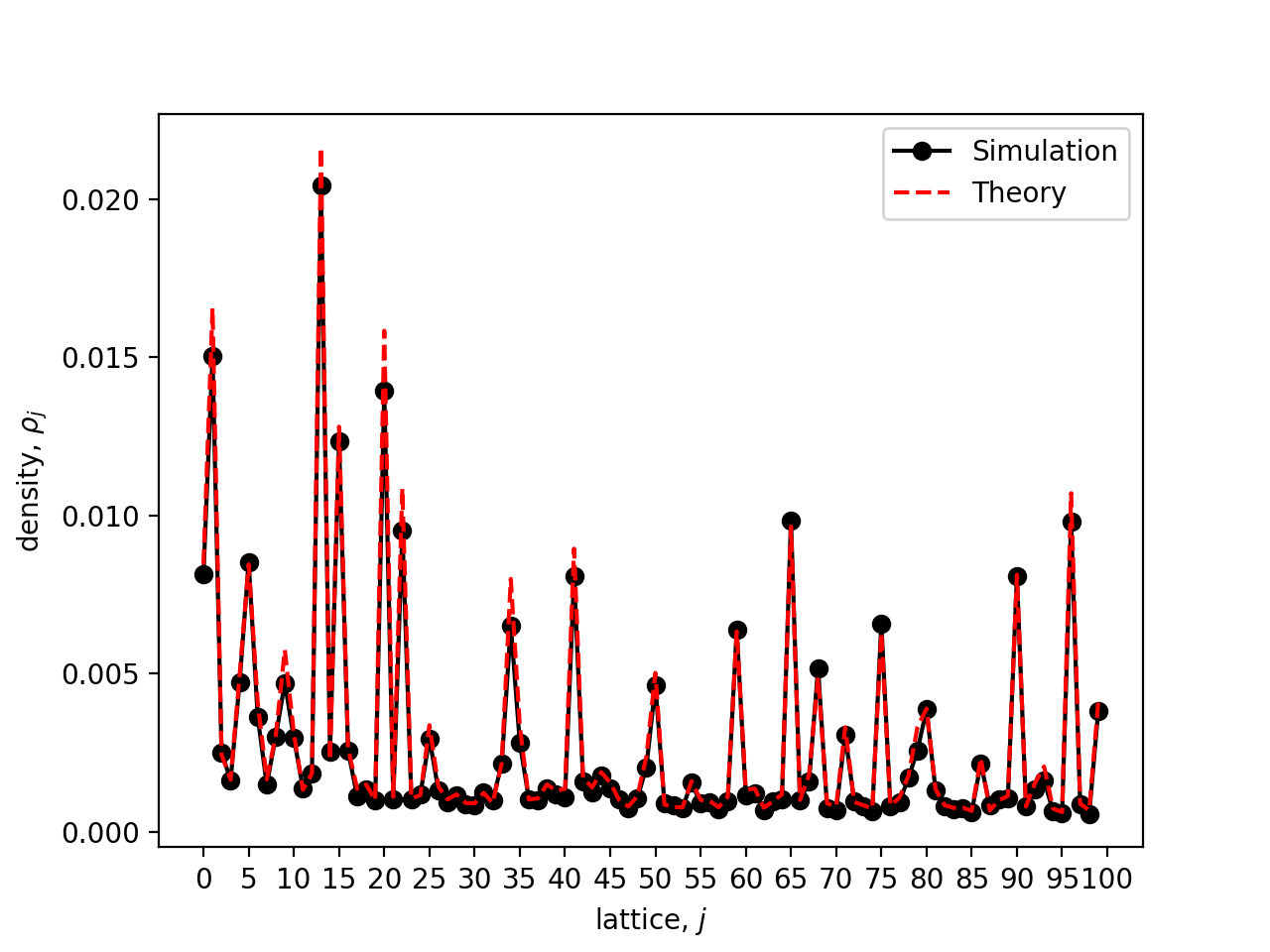}
\includegraphics[scale=.5]{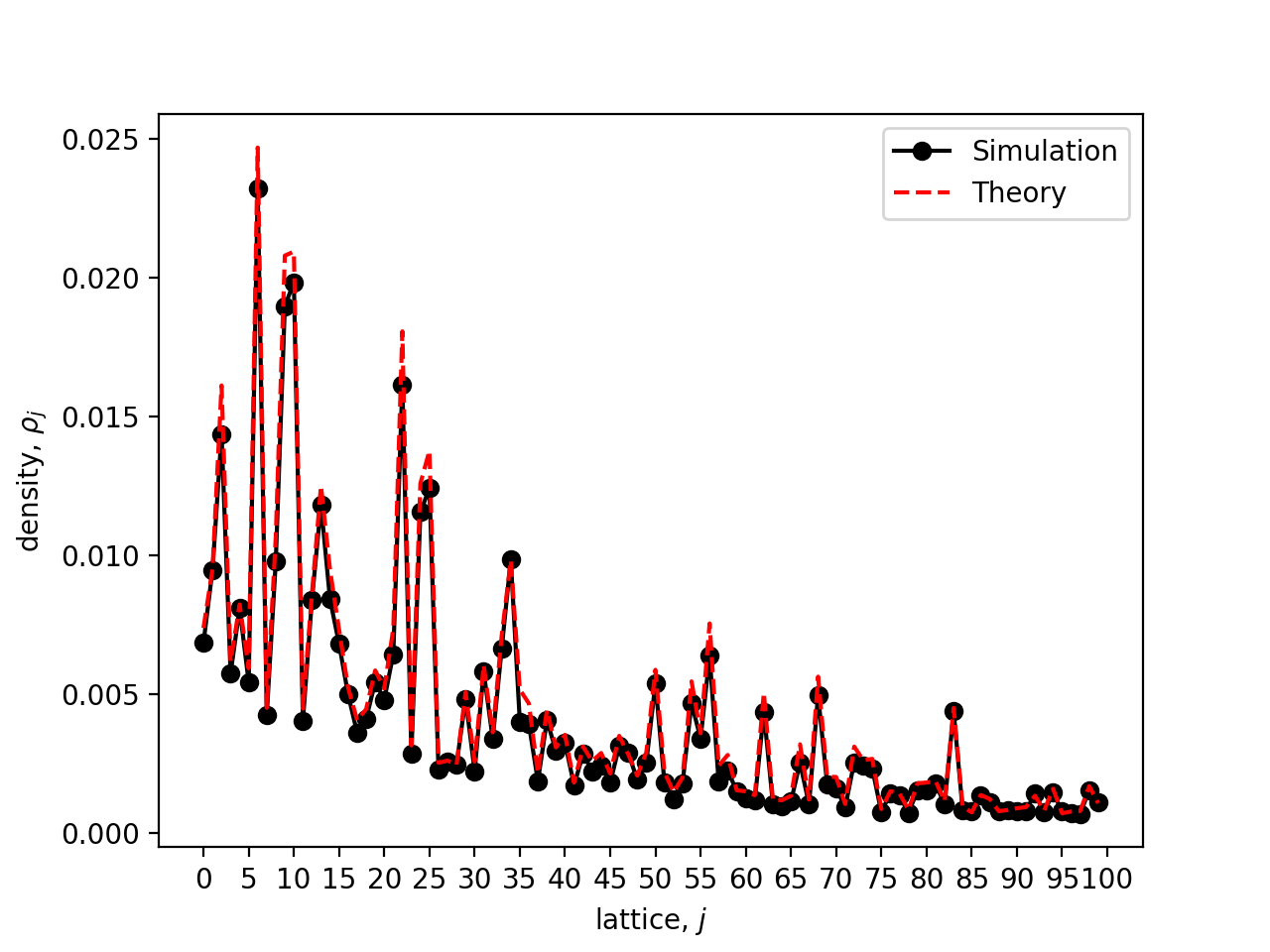}
\caption{(Jumping rates disorder, $0 < \Delta < 1$) Densities, $\rho_j$, plots with the same $\alpha=0.1$, $\beta=30$, $\Omega=0.2$, $\ell=9$, and $N=100$ total lattice sites. ({\bf left}) $\rho_j$ plot with sample sequence drawn from $w_j \in [3,80]$ with $\Delta = 0.96$. ({\bf right}) $\rho_j$ plot with sample sequence drawn from $w_j \in [3,20]$ with $\Delta = 0.85$. The dashed red lines are the Eqn. (\ref{eqn:first:order:solution}) fit to the Monte Carlo simulation data.}
\label{fig:density_2}
\end{figure}

\twocolumngrid
\end{widetext}

\subsubsection{Approximation error distribution}
\begin{figure}
\includegraphics[scale=.5]{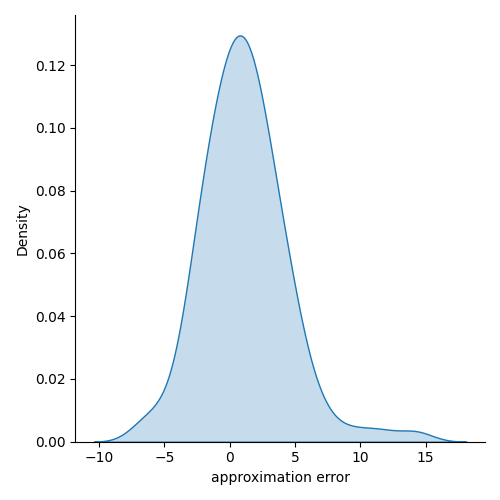}
\caption{Distribution of the particle density approximation error.}
\label{fig:approximate_error_pdf}
\end{figure}
The first order approximation reproduce the exact densities for wide range of the parameter space in the low density phase. We define the relative error
\begin{equation}
\mbox{Approximation-Error}  = \left\{ \frac{\rho_j^{MC} - \rho_j^{approx}}{\rho_j^{MC}}  \right\} \ .
\end{equation}
We plot a sample distribution of the relative errors in percentages (see Fig. \ref{fig:approximate_error_pdf}.

\section{Discussion and conclusions}
We have studied the totally asymmetric exclusion process with quenched jump rates disorder and finite lifetime of the chain. We outlined an alternative approach to efficiently compute chain-age averaged densities from algebraic equations rather than solving relatively computationally expensive differential equations. In addition, we provide a straightforward framework to progressively include correlations by solving computationally inexpensive few sites TASEP systems. Meanwhile, in the biologically relevant Low Density phase, we found that the first leading order asymptotic expansion in the entry rate is accurate for wide range of the system parameters. Plots are shown in Figs (\ref{fig:density_1} and \ref{fig:density_2}) demonstrating the agreement with Monte-Carlo simulation results.

The success of the entry limited approximation is complimented by the fact that the finite degradation rate of the chain favors low density profile downstream (see Fig. \ref{fig:density_profiles0}). Therefore, profiles like that of Fig. (\ref{fig:phase:separation}) won't appear for finite degradation rate $\Omega \neq 0$ and sufficiently long chain.

Even though TASEP is analytically complex problem with rich physics, simple analytic expressions accurate for wide range of parameter space could be found. This is in addition alternative path to including longer range correlations by exploiting the exact solution of few site TASEP. This approach could find applications in highly correlated systems.

\section{Appendix}

\subsection{Entry limited dynamics: Perturbative expansion in $\epsilon$}
We re-write eqn. (\ref{eqn:first:of:marginals:equation}) from the main text to a dimensionless form
\begin{align}
\rho_j = \lambda_{j-1} \rho_{j-1} \left( 1 - \rho_{j} \right)  +  \rho_j \rho_{j+1}  -   \chi_j \rho_j \rho_{j+1}\  , \label{eqn:first:of:marginals:equation:dimensionless:appendix}
\end{align}
where we define dimensionless parameters
\begin{equation}
\lambda_j = \frac{w_j}{ \Omega +  w_{j+1}} \quad \mbox{and} \quad \chi_j  = \frac{\Omega}{\Omega + w_j}  \  .
\end{equation}
Note that $\chi_j$'s are bounded ($0 \leqslant \chi_j \leqslant 1 , \forall \, i$) since $w_j > 0 \  \forall \, i$.

There are two interesting limiting cases: the stable infinite lifetime of the chain, $\Omega \rightarrow 0$, 
$
\lambda_j \rightarrow {w_j}/{w_{j+1}}  \ \mbox{and} \ \chi_j \rightarrow 0  ,
$
and the short-lived chain limit $\Omega \rightarrow \infty$; 
$
\lambda_j \rightarrow 0 \ \mbox{and} \  \chi_j \rightarrow 1 , 
$
where the densities vanish $\rho_j \rightarrow 0, \ \forall j$.

In this entry limited regime, $\lambda_0$ is the small dimensionless parameter of interest (since $\lambda_0 \ll \lambda_j$ for any $\Omega > 0$). Hence, we let $\epsilon \equiv \lambda_0 = \alpha/(\Omega + w_1)$ and then expand the densities in powers of the small parameter $\epsilon$:
\begin{align}
\rho_j & = \epsilon \, \rho_j^{(1)} + \epsilon^2 \, \rho_j^{(2)}  + \cdots  \label{eqn:density:expansion:ansatz}
\end{align}
\emph{First order contribution ($\epsilon^1$):} The first order contribution to the densities are $\rho^{(1)}_1 = 1$ and
\begin{equation}
    \rho^{(1)}_j \  =  \   \prod_{n=1}^{j-1} \left( \frac{w_{n}}{\Omega + w_{n+1}}  \right)  \  ,  \quad  \mbox{for}  \quad  2 \leqslant j  \leqslant N \  .  \label{eqn:first:order:density}
\end{equation}
Therefore, to linear order in the small parameter $\epsilon$, the densities are $\rho_j  \approx   \epsilon  \rho_j^{(1)} $ and
\begin{equation}
\rho_j \  \approx \     \prod_{n=0}^{j-1} \left( \frac{w_{n}}{\Omega + w_{n+1}}  \right)    \  ,   \label{eqn:linear:order}
\end{equation}
for $1\leqslant j \leqslant N$, where we recall that $\epsilon = \alpha/(\Omega + w_1)$ and $w_0 \equiv \alpha$. Therefore, the corresponding particles entry flux is $J_{\text{in}} = \alpha \left( \Omega + w_1 - \alpha \right) /(\Omega + w_1)$.

\emph{Second order contribution ($\epsilon^2$):} The second order contributions accounts for the exclusion effects (and the extended size effects (see the Supporting Information)) where now the densities are
\begin{equation}
\rho_j^{(2)}  = \rho_j^{(1)} \left( \rho_{j+1}^{(1)} -  \sum_{m=1}^{j}  \chi_m \, \rho_{m+1}^{(1)} - \rho_1^{(1)}   \right)  \  , \label{eqn:second:order:density}
\end{equation}
for $1 \leqslant j \leqslant N-1$, while for $j=N$,
\begin{equation}
\rho_N^{(2)}  = \rho_N^{(1)} \left( \rho_{N}^{(1)} -  \sum_{m=1}^{N-1}  \chi_m \, \rho_{m+1}^{(1)} - \rho_1^{(1)}   \right)  \  .
\end{equation}

%
%

\begin{widetext}

\subsection{Few sites exact TASEP solutions}
We present the exact analytic solution of $N=2$ and $N=3$ TASEP systems ($\ell=1$). For brevity we introduce to shorthand
$$
\tau_{\alpha} = \frac{1}{\alpha} \quad , \quad \tau_j = \frac{1}{w_j} \quad \mbox{and} \quad \tau_{\beta} = \frac{1}{\beta}
$$
It should be emphasized here that when using these expressions to approximate the correlations, $\alpha$ and $\beta$ should be interpreted as
\begin{equation}
\alpha \equiv w_{j-1} \rho_{j-1} \quad \mbox{and} \quad \beta \equiv w_{j+S-1} (1-\rho_{j+S})
\end{equation}
where $S$ is the chosen small system size.

\subsubsection{Two (2) sites}
For two sites, we have the distribution 
\begin{equation}
\begin{bmatrix}
\Pt_{00}  \\
\Pt_{01}  \\
\Pt_{10}  \\
\Pt_{11}  
\end{bmatrix}   
= 
\frac{1}{\mathcal{Z}_2}  \begin{bmatrix}
\tau_{alpha}^{2}  \\
\tau_{alpha} \tau_{beta} \\
w_1^{-1} \left( \tau_{alpha} + \tau_{beta} \right) \\
\tau_{beta}^{2}
\end{bmatrix}  \  ,
\end{equation}
where $\mathcal{Z}_2 = \tau_{alpha}^{2} + \tau_{alpha}\tau_{beta} + \tau_1 \left( \tau_{alpha}+\tau_{beta} \right) + \tau_{beta}^{2}$ and the current reads
\begin{equation}
J^{\text{(2-sites)}} \equiv \alpha \Pt_{00} + \alpha \Pt_{01}  = \frac{1}{\mathcal{Z}_2} \left(\tau_{alpha} + \tau_{beta}  \right)   \  .
\end{equation}

\subsubsection{Three (3) sites}
The $N=3$ is a bit more involved but tractable.
\begin{equation}
\begin{bmatrix}
\Pt_{000}  \\
 \\
\Pt_{001}  \\
\\
\Pt_{010}  \\
\\
\Pt_{011}   \\
\\
\Pt_{100} \\
\\
\Pt_{101} \\
\\
\Pt_{110} \\
\\
\Pt_{111}
\end{bmatrix}   
= 
\frac{1}{\mathcal{Z}_3}  \begin{bmatrix}
\tau_{\alpha}^{3} \Big( \tau_{\alpha} \tau_{\beta} + \tau_{\alpha} \tau_1 + \tau_{\beta} \tau_1 \Big)  \\
\tau_{\alpha}^{2} \tau_{\beta} \Big( \tau_{\alpha} \tau_{\beta} + \tau_{\alpha} \tau_1 + \tau_{\beta} \tau_1 \Big)  \\
\tau_{\alpha} \tau_2 \Big( \tau_{\alpha} + \tau_{\beta}  \Big) \Big( \tau_{\alpha} \tau_{\beta} + \tau_{\alpha} \tau_1 + \tau_{\beta} \tau_1 \Big) \\
\tau_{\alpha} \tau_{\beta}^2 \Big( \tau_{\alpha} \tau_{\beta} + \tau_{\alpha} \tau_2 + \tau_{\beta} \tau_2  \Big)  \\ 
\tau_1^2 \Big( \tau_{\alpha} + \tau_{\beta} \Big) \Big(   \tau_{\alpha} \tau_{\beta} + \tau_{\alpha} \tau_2 + \tau_{\beta} \tau_2 \Big) + \tau_{\alpha}^2 \tau_1 \Big(  \tau_{\alpha} \tau_{\beta} + \tau_{\alpha} \tau_1 + \tau_{\beta} \tau_1 \Big) \\
\tau_{\beta} \tau_1 \Big( \tau_{\alpha} + \tau_{\beta}  \Big) \Big( \tau_{\alpha} \tau_{\beta} + \tau_{\alpha} \tau_2 + \tau_{\beta} \tau_2 \Big) \\
\tau_2^2 \Big( \tau_{\alpha} + \tau_{\beta} \Big) \Big(   \tau_{\alpha} \tau_{\beta} + \tau_{\alpha} \tau_1 + \tau_{\beta} \tau_1 \Big) + \tau_{\beta}^2 \tau_2 \Big(  \tau_{\alpha} \tau_{\beta} + \tau_{\alpha} \tau_2 + \tau_{\beta} \tau_2 \Big) \\
\tau_{\beta}^{3} \Big( \tau_{\alpha} \tau_{\beta} + \tau_{\alpha} \tau_2 + \tau_{\beta} \tau_2 \Big) 
\end{bmatrix}
\end{equation}
where the partition function 
\begin{align}
\mathcal{Z}_3 & = \Big( \tau_{\alpha}^2 \left[ \tau_{\alpha} + \tau_{\beta} + \tau_1 \right] + \tau_2 \left( \tau_{\alpha} + \tau_{\beta} \right) \left( \tau_{\alpha} + \tau_2 \right) \Big)  \Big( \tau_{\alpha} \tau_{\beta} + \tau_{\alpha} \tau_1 + \tau_{\beta} \tau_1 \Big) \nonumber  \\
& \qquad \qquad \qquad \qquad  +  \Big( \tau_{\beta}^2 \left[ \tau_{\alpha} + \tau_{\beta} + \tau_2 \right] + \tau_1 \left( \tau_{\alpha} + \tau_{\beta} \right) \left( \tau_{\alpha} + \tau_1 \right) \Big) \Big( \tau_{\alpha} \tau_{\beta} + \tau_{\alpha} \tau_2 + \tau_{\beta} \tau_2 \Big)  \ .
\end{align}
The current then reads
\begin{equation}
J^{\text{(3-sites)}} = \frac{1}{\mathcal{Z}_3} \left( \tau_{\alpha} + \tau_{\beta}  \right)  \left( \tau_{\alpha} + \tau_2  \right)  \Big( \tau_{\alpha} \tau_{\beta} + \tau_{\alpha} \tau_1 + \tau_{\beta} \tau_1 \Big)  + \frac{1}{\mathcal{Z}_3} \tau_{\beta}^2  \Big( \tau_{\alpha} \tau_{\beta} + \tau_{\alpha} \tau_2 + \tau_{\beta} \tau_2 \Big)  \  .
\end{equation}
%
%
%
\end{widetext}

\bibliographystyle{acm}
\bibliography{inhomogeneous_tasep23.bib}

\end{document}